\documentclass[12pt]{article}

\usepackage[totalwidth=483truept,totalheight=623truept]{geometry}
\usepackage{amsfonts,latexsym,amssymb,amsmath,graphicx,accents,slashed,subfigure}
\usepackage{dsfont}
\usepackage[T1]{fontenc}
\usepackage[hidelinks]{hyperref}

\global\arraycolsep=1truept

\numberwithin{equation}{section}

\begin{document}

\phantom{C}

\vskip2truecm

\begin{center}
{\huge \textbf{Perturbative Unitarity Of}}

\vskip.4truecm

{\huge \textbf{Lee-Wick Quantum Field Theory }}

\vskip1.5truecm

\textsl{Damiano Anselmi\footnote{%
damiano.anselmi@unipi.it} and Marco Piva\footnote{%
marco.piva@df.unipi.it}}

\vskip .2truecm

\textit{Dipartimento di Fisica \textquotedblleft Enrico
Fermi\textquotedblright , Universit\`{a} di Pisa, }

\textit{Largo B. Pontecorvo 3, 56127 Pisa, Italy}

\textit{and INFN, Sezione di Pisa,}

\textit{Largo B. Pontecorvo 3, 56127 Pisa, Italy}

\vskip2truecm

\textbf{Abstract}
\end{center}

We study the perturbative unitarity of the Lee-Wick models, formulated as
nonanalytically Wick rotated Euclidean theories. The complex energy plane is
divided into disconnected regions and the values of a loop integral in the
various regions are related to one another by a nonanalytic procedure. We
show that the one-loop diagrams satisfy the expected, unitary cutting
equations in each region: only the physical degrees of freedom propagate
through the cuts. The goal can be achieved by working in suitable subsets of
each region and proving that the cutting equations can be analytically
continued as a whole. We make explicit calculations in the cases of the
bubble and triangle diagrams and address the generality of our approach. We
also show that the same higher-derivative models violate unitarity if they
are formulated directly in Minkowski spacetime.

\vfill\eject

\section{Introduction}

\setcounter{equation}{0}

The nonrenormalizability of the Hilbert-Einstein Lagrangian \cite%
{thooftveltman} teaches us that, if we want to solve the problem of quantum
gravity, we have to explore new sectors of quantum field theory and maybe
relax some assumptions we are accustomed to. In this respect, an interesting
subsector of quantum field theory is represented by the local,
higher-derivative theories, because there is still a possibility that the
search for a consistent theory of quantum gravity might lead there.

However, the formulation of higher-derivative theories turns out to be less
trivial than expected. For example, when the free propagators have complex
poles, the theories cannot be consistently defined in Minkowski spacetime 
\cite{ugo}, in general, because they generate nonlocal, non-Hermitian
divergences, which cannot be subtracted away without destroying the nature
of the theory itself.

The Lee-Wick (LW) models \cite{LW} are a special subclass of local,
higher-derivative theories, which have the possibility of reconciling
renormalizability and unitarity. The propagators contain complex conjugate
pairs of extra poles, which we call \textit{LW poles}, besides the poles
corresponding to the physical degrees of freedom and the degrees of freedom
introduced by the gauge fixing (which are those propagated by the temporal
and the longitudinal components of the gauge fields, as well as the
Faddeev-Popov ghosts). The Lee-Wick models are claimed to lead to a
perturbatively unitary $S$ matrix \cite{LW,LWqed,CLOP}. Because of their
unusual features, their formulation has been the object of several
investigations. Like all higher-derivative theories, they violate
microcausality. Nakanishi \cite{nakanishi} showed that if the loop space
momenta are integrated on their natural, real values, as Lee initially
seemed to suggest \cite{lee}, Lorentz invariance is violated. Cutkosky 
\textit{et al}. (CLOP) showed \cite{CLOP} that the $S$ matrix is not
analytic when pairs of LW poles pinch the integration path of the energy.
They proposed to treat such a pinching, which we call \textit{LW pinching},
by means of a limiting procedure, known as \textit{CLOP prescription}. Among
other things, the CLOP\ prescription removes the problems found by
Nakanishi. In simple diagrams, it gives an unambiguous, Lorentz invariant
and unitary result, as confirmed by the calculations of Grinstein \textit{et
al}. \cite{grinstein}. However, it seems a bit artificial, since it cannot
be incorporated into a Lagrangian and ambiguities are still present. For a
while it was thought that such ambiguities survived only at high orders
diagrams \cite{CLOP}, but recently it has been shown that they are present
also at one loop \cite{LWformulation}.

These pieces of information need to be clarified and properly assessed. To
answer some of the open questions, a new formulation of the Lee-Wick models
has been recently proposed \cite{LWformulation}, by viewing them as \textit{%
nonanalytically Wick rotated Euclidean higher-derivative theories}.

Since the Minkowski formulation is not viable \cite{ugo}, we have no choice
but start from the Euclidean version of the higher-derivative theories.
However, the Wick rotation turns out to be nonanalytic, because of the LW\
pinching, to the extent that the complex energy plane is divided into
disjoint regions $\mathcal{A}_{i}$ of analyticity. The Lorentz violation is
avoided by working in a generic Lorentz frame, with generic external
momenta, and deforming the integration domain of the space loop momenta to
complex values in a suitable way. It turns out that the models are
intrinsically equipped with all that is necessary to define them properly.
There is no need of the CLOP prescription, or any other prescription, to
handle the pinching of the LW poles. Moreover, the CLOP prescription leads
to physical results that are ambiguous, even in a simple case such as the
bubble diagram with different physical masses \cite{LWformulation}.
Therefore, the \textit{ad hoc} prescriptions should be dropped.

Because the Lee-Wick models have been reformulated anew, and the new
formulation leads to predictions that are quantitatively different from
those of the previous approaches, it is compulsory to investigate
perturbative unitarity in the new formulation. Writing the $S$ matrix as $%
S=1+iT$, the unitarity relation $SS^{\dagger }=1$ is equivalent to $%
T-T^{\dag }=iTT^{\dag }$. This identity can be expressed diagrammatically by
means of the so-called cutting equations \cite{unitarity}, which relate the
discontinuity of an amplitude to the sum of cut diagrams (see also \cite%
{unitaritymio} for a recent extension and \cite{ACE} for an algebraic
reformulation). The cut diagrams are built with cut propagators and shadowed
vertices, in addition to the usual propagators and vertices. In this paper,
we study the cutting equations in the one-loop bubble and triangle diagrams
explicitly, but the procedure can be extended to all the one-loop diagrams.

The cutting equations must be derived within the formulation of the models
as nonanalytically Wick rotated Euclidean theories. To achieve this goal, we
show that it is possible to derive the cutting equations in suitable subsets 
$\mathcal{O}_{i}$ of the analytic regions $\mathcal{A}_{i}$ and extend their
validity to the whole $\mathcal{A}_{i}$ by analytic continuation. The
analytic continuation of the cut diagrams is something that also requires
some attention, because it is not discussed in the available literature.

The results we find confirm that the cutting equations of the LW models are
consistent with perturbative unitarity. The contributions of the poles of
each LW pair mutually cancel, so only the physical degrees of freedom
propagate through the cuts.

Our findings also suggest that the cancellation mechanism, which is encoded
in formula (\ref{crucial}), is a general property of all diagrams. While the
bubble diagram is too special to argue in favor of general properties, the
derivation of the cutting equations for the triangle diagram is sufficiently
general to be applied to all the one-loop diagrams. The generalization to
diagrams with more loop is less direct, but it appears to be mostly a
technical matter, which is why we believe that our results can be the
starting point to derive a proof of perturbative unitarity to all orders.

Finally, to emphasize the importance of the nonanalytic Wick rotation, we
show that the same higher-derivative models do violate unitarity when they
are defined directly in Minkowski spacetime.

The LW models are important not only theoretically, but also because they
may have phenomenological applications. Among those that have been
considered in the literature, we mention extensions of QED \cite{LWqed},
physics beyond the standard model \cite{LWstandardM} and grand unified
theories \cite{LWunification}, as well as the search for a consistent theory
of quantum gravity \cite{LWgrav,shapiromodesto}. In ref. \cite{LWformulation}
it was also noted that the unusual behaviors of the physical amplitudes, due
to the violations of analyticity, may have important phenomenological
consequences, for example allow us to measure some key physical constants of
the LW models, such as the scales associated with the higher-derivative
terms.

The paper is organized as follows. In section \ref{LWE} we recall the
formulation of the models. In section \ref{anacut} we study the analytic
continuation of the cut diagrams. In section \ref{LWun} we reconsider the
bubble diagram in standard theories and derive its cutting equations in a
setting that is sufficiently general to ease out the extension to the LW
models. In section \ref{LWbubble}, we derive the cutting equations of the
bubble diagram in the LW models and show that only the physical degrees of
freedom propagate through the cuts. In section \ref{LWuntriangle}, we do the
same for the triangle diagram. In section \ref{nontrinum}, we extend our
results to Feynman diagrams with nontrivial numerators and comment on the
validity of our arguments in arbitrary diagrams. In section \ref{Minkviol},
we show that the higher-derivative theories of the LW class, if defined
directly in Minkowski spacetime, do violate unitarity. Section \ref{conclu}
contains our conclusions.

\section{Lee-Wick models as nonanalytically Wick rotated Euclidean theories}

\setcounter{equation}{0}

\label{LWE}

In this section we recall how the LW\ models are formulated as
nonanalytically Wick rotated Euclidean theories \cite{LWformulation}. For
concreteness, it may be useful to have a specific theory in mind, such as
the massive Lee-Wick $\varphi ^{4}$ theory in four dimensions described by
the Lagrangian 
\begin{equation}
\mathcal{L}=\frac{1}{2}(\partial _{\mu }\varphi )\left( 1+\frac{\square ^{2}%
}{M^{4}}\right) (\partial ^{\mu }\varphi )-\frac{1}{2}m^{2}\varphi \left( 1+%
\frac{\square ^{2}}{M^{4}}\right) \varphi -\frac{\lambda }{4!}\varphi ^{4},
\label{lfi4}
\end{equation}%
whose free propagator reads in momentum space 
\begin{equation}
iD(p^{2},m^{2},\epsilon )=\frac{iM^{4}}{(p^{2}-m^{2}+i\epsilon
)((p^{2})^{2}+M^{4})}.  \label{propa}
\end{equation}%
More general propagators and more diverse theories can be considered, but
they do not change the sense of the discussion that follows. In the limit $%
M\rightarrow \infty $, (\ref{propa}) returns the standard propagator, while
at $M<\infty $ extra poles, which we call LW poles, are present besides the
standard ones. The LW\ poles come in complex conjugate pairs, which we call 
\textit{LW\ pairs}.

\begin{figure}[t]
\begin{center}
\includegraphics[width=8cm]{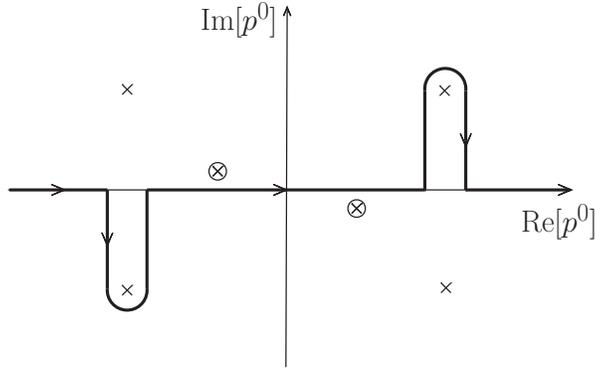}
\end{center}
\caption{Integration path given by the Wick rotation}
\label{wick}
\end{figure}

\begin{figure}[b]
\begin{center}
\includegraphics[width=4truecm]{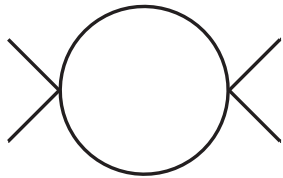}
\end{center}
\caption{Bubble diagram}
\label{bubble}
\end{figure}

\begin{figure}[t]
\begin{center}
\includegraphics[width=11cm]{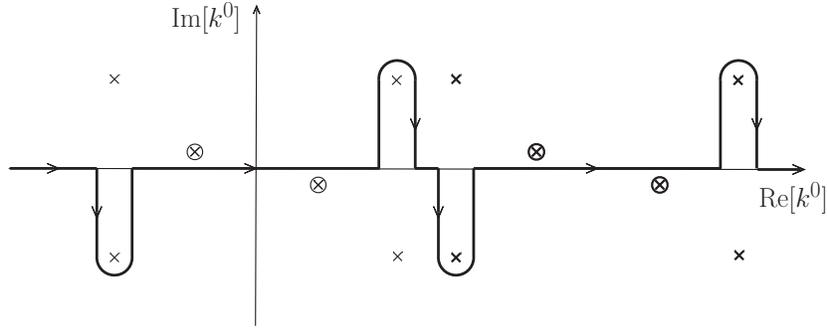}
\end{center}
\caption{Integration path of the bubble diagram}
\label{WickBub2}
\end{figure}

The Wick rotation is simple for a single propagator. When the imaginary axis
is rotated to the real one, we get the integration path of fig. \ref{wick},
where the encircled crosses denote the standard poles and the non encircled
crosses denote the LW poles. In generic Feynman diagrams, where more
propagators are present, the Wick rotation is less trivial. Let us consider,
for example, the bubble diagram, fig. \ref{bubble}. The loop integral is
proportional to 
\begin{equation}
\mathcal{J}(p)=\int \frac{\mathrm{d}^{D}k}{(2\pi )^{D}}D(k^{2},m_{1}^{2},%
\epsilon _{1})D((p-k)^{2},m_{2}^{2},\epsilon _{2}).  \label{bubd}
\end{equation}%
For the sake of generality, we take different masses $m_{1}$, $m_{2}$ and
different infinitesimal widths $\epsilon _{1}$, $\epsilon _{2}$. When we
vary the external momentum $p$, the poles of the first propagator are fixed,
while those of the second propagator move on the complex $k^{0}$ plane.
Assuming for simplicity that the external space momentum $\mathbf{p}$
vanishes and taking $p^{0}$ real, the Wick rotation gives the integration
path of fig. \ref{WickBub2}.

We see that the left LW\ pair of a propagator is always above the
integration path, while the right LW pair is always below. This property
holds in arbitrary diagrams. A pinching, which we call \textit{LW\ pinching}%
, occurs when the left (right) LW pair of the first propagator hits the
right (left) LW\ pair of the second propagator. The threshold of this
pinching is $p^{2}=2M^{2}$. With complex $p^{0}$, other types of LW\
pinchings occur: the bottom LW\ pole of the left LW\ pair of the first
propagator can hit the top LW\ pole of the right LW\ pair of the second
propagator, and so on. The thresholds of these pinchings are $p^{2}=\pm
4iM^{2}$. Several such situations are symmetric to one another, so it
suffices to study a single representative of each symmetric subset.

The threshold associated with a LW\ pinching will be called \textit{LW\
threshold}. We anticipate that the LW\ thresholds are not associated with
discontinuities of the amplitudes, in agreement with unitarity. However,
they are associated with nonanalytic behaviors of the amplitudes.

We focus on the pure LW pinching, which involves two LW poles, because at
one loop it is the only LW pinching that has thresholds on the real axis.
The mixed LW pinching, which occurs between a LW pole and a standard pole,
needs complex external momenta $p$ and its thresholds are far away from the
real axis.

\begin{figure}[b]
\begin{center}
\includegraphics[width=8truecm]{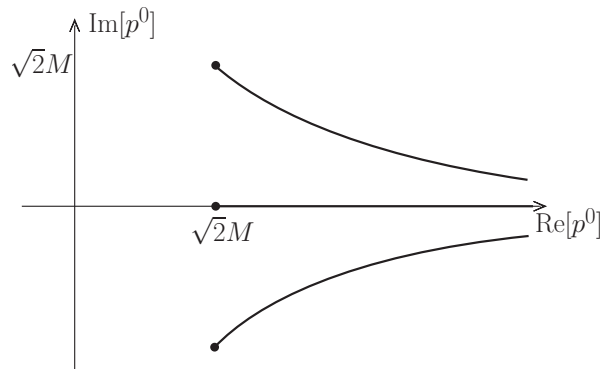}
\end{center}
\caption{LW thresholds and LW branch cuts at $\mathbf{p}=0$}
\label{compl}
\end{figure}

To evaluate $\mathcal{J}(p)$, we first integrate over the loop energy $k^{0}$
by means of the residue theorem, after which we remain with the integral on
the loop space momentum $\mathbf{k}$. At $\mathbf{p}=0$, if we keep the $%
\mathbf{k}$ integration domain rigid, i.e. integrate $\mathbf{k}$ on its
natural, real values, the positions of the LW thresholds and the LW branch
cuts as functions of the external (complex) energy $p^{0}$ are those shown
in fig. \ref{compl}, plus their reflections with respect to the imaginary
axis. In particular, the branch cut on the real axis is made by the
solutions of the pinching condition%
\begin{equation*}
p^{0}=\sqrt{\mathbf{k}^{2}+iM^{2}}+\sqrt{\mathbf{k}^{2}-iM^{2}}.
\end{equation*}%
When $p^{0}$ crosses one of the curves shown there, a pole of the integrand
crosses the $\mathbf{k}$ integration domain. The cuts can be analytically
deformed by deforming the $\mathbf{k}$ integration domain before it is
crossed by the pole, so as to prevent the crossing from actually occurring.

\begin{figure}[t]
\begin{center}
\includegraphics[width=8truecm]{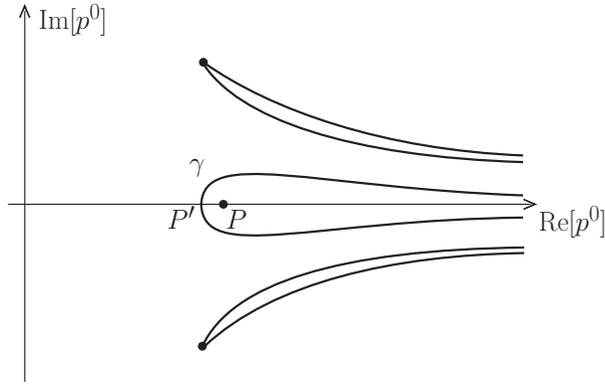}
\end{center}
\caption{Areas of LW pinching at $\mathbf{p}\neq 0$}
\label{compl2}
\end{figure}

Something interesting happens at nonvanishing, real $\mathbf{p}$. Keeping
the $\mathbf{k}$ integration domain rigid again, we find that each cut of
figure \ref{compl} enlarges into the regions $\mathcal{\tilde{A}}_{i}$ shown
in fig. \ref{compl2}. We denote the main region, i.e. the one that contains
the imaginary axis, by $\mathcal{\tilde{A}}_{0}$. There, the Wick rotation
is analytic, because no LW pinching occurs. The curve $\gamma $ is the
boundary of a different region, which we denote by $\mathcal{\tilde{A}}_{P}$%
, which contains the positive real axis above the threshold $p^{2}=2M^{2}$,
located in the point $P$. The points of $\mathcal{\tilde{A}}_{P}$ are the
solutions of the pinching condition%
\begin{equation}
p^{0}=\sqrt{\mathbf{k}^{2}+iM^{2}}+\sqrt{(\mathbf{k-p})^{2}-iM^{2}}.
\label{pinchcond}
\end{equation}%
Note that $\gamma $ does not intersect the real axis in $P$, but in another
point $P^{\prime }$, located below the threshold. Working out the
coordinates of $P$ and $P^{\prime }$, we find%
\begin{equation}
P:\qquad p^{0}=\sqrt{2M^{2}+\mathbf{p}^{2}}\equiv E_{P},\qquad \qquad
P^{\prime }:\qquad p^{0}=\sqrt{\frac{\mathbf{p}^{2}}{2}+\sqrt{\frac{(\mathbf{%
p}^{2})^{2}}{4}+4M^{4}}}\equiv E_{P^{\prime }},  \label{ab}
\end{equation}%
which satisfy $\sqrt{2}M<E_{P^{\prime }}<E_{P}$, $E_{P}-E_{P^{\prime }}<%
\sqrt{2}M$.

Since the location of $P^{\prime }$ has no Lorentz invariant meaning,
Lorentz invariance appears to be violated. Recall that fig. \ref{compl2} is
derived by keeping the loop space momentum $\mathbf{k}$ real. However, the
integration path of fig. \ref{wick} shows that the loop energy is not
everywhere real, so Lorentz invariance implies the loop momentum cannot be
everywhere real. To recover Lorentz invariance, the $\mathbf{k}$ integration
domain must be deformed to include complex values. Moreover, the deformation
must turn the surfaces of fig. \ref{compl2} into Lorentz invariant lines
(i.e. solutions of Lorentz invariant conditions), similar to those of fig. %
\ref{compl}. In particular, it must turn the region $\mathcal{\tilde{A}}_{P}$
into the half line of the real axis that goes from the point $P$ to $+\infty 
$, which we denote by $\mathcal{O}_{P}$. Indeed, $\mathcal{O}_{P}$ is
Lorentz invariant, while any extended region is not.

It can be argued \cite{LWformulation} that the domain deformation just
described restores both Lorentz invariance and analyticity above the LW
threshold. To give more details on this, let us write the propagator (\ref%
{propa}) in the equivalent form%
\begin{equation}
iD_{0}(p^{2},m^{2},\epsilon )+iD_{\text{LW}}(p^{2},m^{2}),  \label{deLW}
\end{equation}%
where 
\begin{equation*}
D_{0}(p^{2},m^{2},\epsilon )=\frac{M^{4}}{M^{4}+m^{4}}\frac{1}{%
p^{2}-m^{2}+i\epsilon },\qquad D_{\text{LW}}(p^{2},m^{2})=-\frac{M^{4}}{%
M^{4}+m^{4}}\frac{p^{2}+m^{2}}{(p^{2})^{2}+M^{4}}.
\end{equation*}%
We can use this decomposition to separate the contributions of the physical
poles from the ones of the LW\ poles in every diagram. Then, we focus on the
contributions that involve LW poles. For example, in the bubble diagram we
take 
\begin{equation}
\mathcal{J}_{\text{LW}}(p)=\int \frac{\mathrm{d}^{D}k}{(2\pi )^{D}}D_{\text{%
LW}}(k^{2},m_{1}^{2})D_{\text{LW}}((k-p)^{2},m_{2}^{2}).  \label{JLW}
\end{equation}

\begin{figure}[t]
\begin{center}
\includegraphics[width=10truecm]{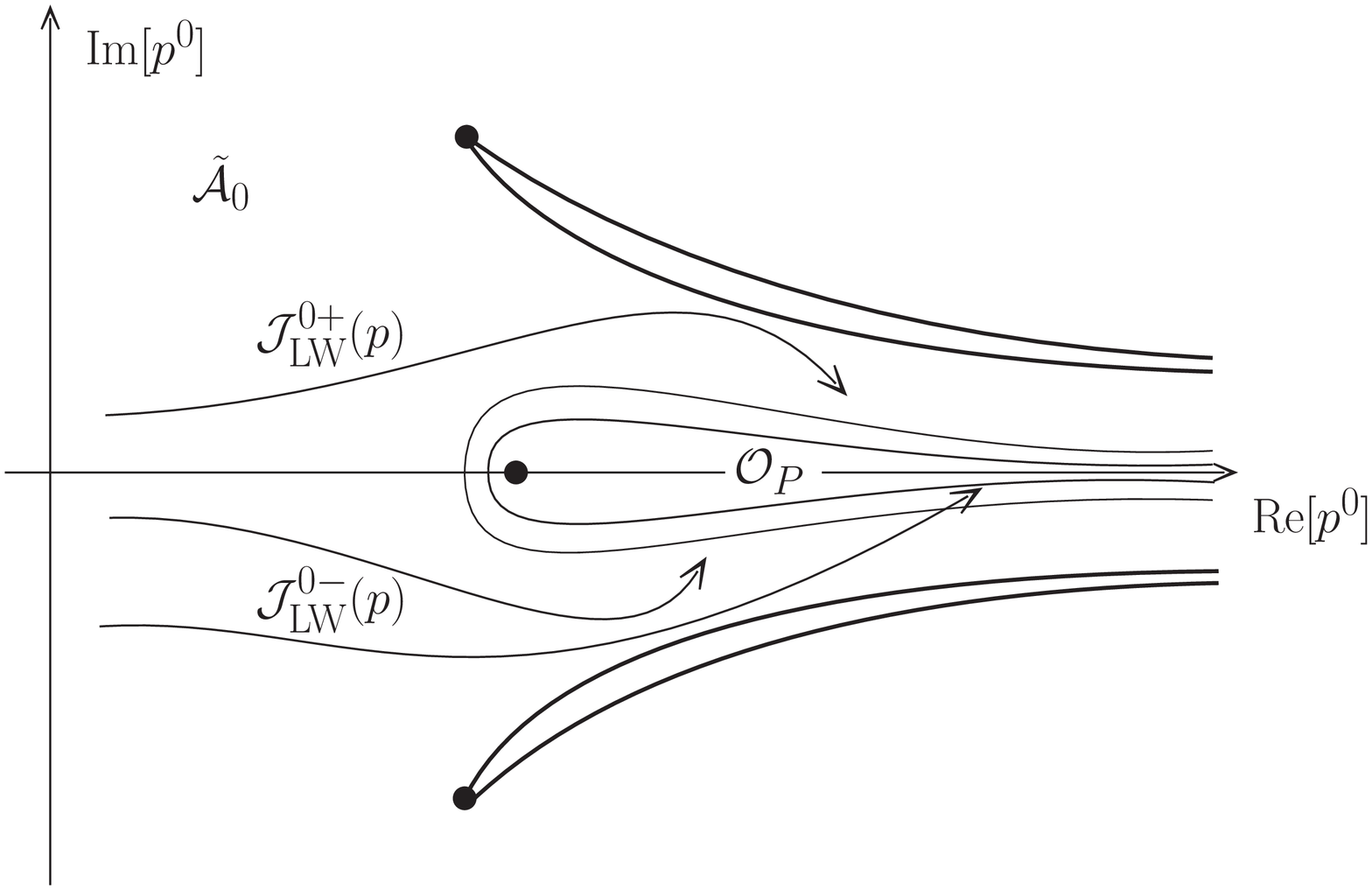}
\end{center}
\caption{Definitions of $\mathcal{J}_{\text{LW}}^{0+}(p)$ and $\mathcal{J}_{%
\text{LW}}^{0-}(p)$}
\label{complJW}
\end{figure}

\begin{figure}[b]
\begin{center}
\includegraphics[width=8truecm]{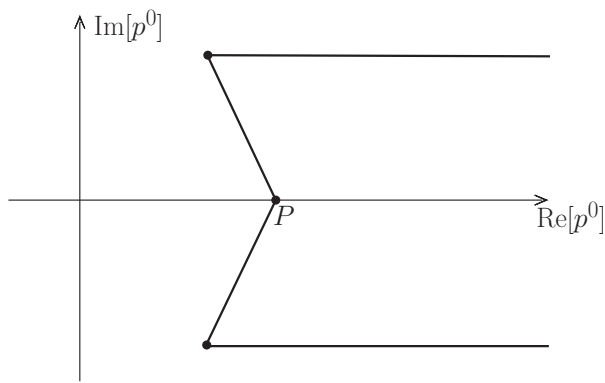}
\end{center}
\caption{Analytic regions}
\label{compl3}
\end{figure}

The function $\mathcal{J}_{\text{LW}}(p)$ is analytic and Lorentz invariant
in the main region $\mathcal{\tilde{A}}_{0}$, because the Wick rotation is
analytic there. In $\mathcal{O}_{P}$ (which means on the real axis above $P$%
) the domain deformation described above leads to the result \cite%
{LWformulation}%
\begin{equation}
\mathcal{J}_{\text{LW}}(p)=\frac{1}{2}\left[ \mathcal{J}_{\text{LW}}^{0+}(p)+%
\mathcal{J}_{\text{LW}}^{0-}(p)\right] ,  \label{jp}
\end{equation}%
where the functions $\mathcal{J}_{\text{LW}}^{0\pm }(p)$ are obtained by
analytically continuing $\mathcal{J}_{\text{LW}}(p)$ from $\mathcal{\tilde{A}%
}_{0}$ to $\mathcal{O}_{P}$ from the half plane Im$[p^{0}]>0$ or from the
half plane Im$[p^{0}]<0$, respectively, as illustrated in fig. \ref{complJW}%
. The continuations can be stretched to neighborhoods of $\mathcal{O}_{P}$
above $P$, to eventually cover an extended region $\mathcal{A}_{P}$ such as
the one shown in fig. \ref{compl3}.

In the end, the complex plane is divided into disjoint regions $\mathcal{A}%
_{i}$ of analyticy. We call $\mathcal{A}_{0}$ the analytic region that
contains the imaginary axis. The function $\mathcal{J}_{\text{LW}}(p)$ is
analytic in each region, but not on the entire complex plane. Formula (\ref%
{jp}) relates the value of the function in $\mathcal{A}_{P}$ to the value of
the function in $\mathcal{A}_{0}$. In particular, it ensures Lorentz
invariance and analyticity in $\mathcal{A}_{P}$ thanks to the Lorentz
invariance and analyticity in $\mathcal{\tilde{A}}_{0}$.

We stress again that the amplitudes must be evaluated at generic external
momenta and in a generic Lorentz frame, because special Lorentz frames may
squeeze some regions $\mathcal{\tilde{A}}_{i}$ into curves $\Gamma _{i}$.
For example, in the center of mass frame $\mathbf{p}=0$, the region $%
\mathcal{\tilde{A}}_{P}$ of the bubble diagram is squeezed onto $\mathcal{O}%
_{P}$. The value of the amplitude in $\mathcal{O}_{P}$ is ill defined at $%
\mathbf{p}=0$, but can be worked out at $\mathbf{p}\neq 0$, where $\mathcal{%
\tilde{A}}_{P}$ is extended, by means of the deformation procedure explained
above. Note that the deformation also squeezes $\mathcal{\tilde{A}}_{P}$
onto $\mathcal{O}_{P}$, but that happens when the amplitude is evaluated
inside of it, not before.

The integrand of $\mathcal{J}(p)$ is singular where the LW pinching occurs,
but the singularity is integrable. Specifically, focus on the intersection $%
\mathcal{O}_{P}$ between $\mathcal{\tilde{A}}_{P}$ and the real axis. The
pinching involves the left LW\ pair of one propagator and the right LW pair
of the other propagator. For $\mathbf{p}$ small the integral around the
pinching of the top LW poles is proportional to \cite{LWformulation}%
\begin{equation}
\frac{\mathrm{d}\tau \mathrm{d}u}{\tau -iC|\mathbf{p|}u},  \label{prescr}
\end{equation}%
where $C$ is a positive, $\mathbf{p}$-independent constant, $u=\cos \theta $%
, $\theta $ being the angle between the vectors $\mathbf{p}$ and $\mathbf{k}$%
, and $\tau $ parametrizes the fluctuation of $|\mathbf{k}|$ around the
value it has at the singularity, which is $\sqrt{(p^{0})^{4}-4M^{4}}%
/(2|p^{0}|)$. The pinching of the bottom LW poles is described by flipping
the sign in front of $iC$.

We see that, basically, a nonvanishing $|\mathbf{p|}$ provides the
prescription for handling the integral. The limit $|\mathbf{p|}\rightarrow 0$
can be evaluated explicitly, because it squeezes $\mathcal{\tilde{A}}_{P}$
onto $\mathcal{O}_{P}$ bypassing the domain deformation. The result is%
\begin{equation}
\mathrm{d}\tau \mathrm{d}u\left[ \mathcal{P}\frac{1}{\tau }+i\pi \mathrm{sgn}%
(u)\delta (\tau )\right] \rightarrow \mathrm{d}\tau \hspace{0.01in}\mathcal{P%
}\frac{1}{\tau },  \label{ter}
\end{equation}%
where $\mathcal{P}$ denotes the principal value and \textquotedblleft
sgn\textquotedblright\ is the sign function. In the last step we have
performed the $u$ integration, which is trivial because the integrand of $%
\mathcal{J}(p)$ is $u$ independent at $|\mathbf{p|}=0$.

Let us describe what happens in more complicated diagrams. At one loop, the
LW pinchings are similar to those of the bubble diagram. They are still
described by fig. \ref{WickBub2} and occur between the LW\ poles of any pair
of propagators. As before, the LW thresholds on the real axis are given by
the formula $p^{2}=2M^{2}$, where now $p$ is any sum of external (incoming)
momenta. In section \ref{LWuntriangle} the triangle diagram is studied in
detail. With more loops, the LW thresholds can involve both LW poles and
physical poles. However, apart from minor differences, the arguments and
properties outlined above --- such as the recovery of analyticity and
Lorentz invariance by means of the domain deformation, the behavior (\ref%
{prescr}) of the integral around the potential singularity due to the LW
pinching, as well as formula (\ref{jp}) --- are still expected to hold,
because their essential features are not related to the specific diagrams we
have considered. More details on this can be found in section \ref{nontrinum}%
.

\section{Analytic continuation of the cut diagrams}

\setcounter{equation}{0}

\label{anacut}

In this section, we explain how to analytically continue the cutting
equations for the study perturbative unitarity. Due to the domain
deformation explained in the previous section, we have to include complex
values of the loop space momenta $\mathbf{k}$. Nevertheless, the
contributions to the cutting equations due to the poles of the same LW pair
still compensate each other. This result is ensured by the key formula (\ref%
{crucial}). That formula only holds at $\epsilon =0$, where standard regions
of the complex plane are squeezed to the real axis (see below). We have to
clarify how to work at nonzero $\epsilon $ and when exactly the limit $%
\epsilon \rightarrow 0$ must be taken, if before or after the domain
deformation.

We first discuss related issues in standard theories, then move to the LW
models.

\begin{figure}[b]
\begin{center}
\includegraphics[width=14truecm]{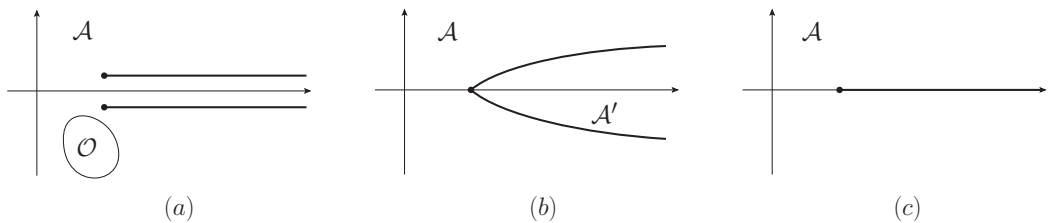}
\end{center}
\caption{Analytic regions of the standard bubble diagram}
\label{standardregios}
\end{figure}

When $\epsilon \rightarrow 0$, the standard pinching takes place. Consider,
for example, the cut version of the standard bubble diagram of a massive
field of mass $m$. The branch points are $p^{0}=\pm 2m$ and the cuts are $%
p^{0}\geqslant 2m$ and $p^{0}\leqslant -2m$, located on the real axis, where 
$p^{0}$ denotes the external energy and $\mathbf{p}$ is assumed to vanish.
Those cuts are squeezed regions, one of which is shown in fig. \ref%
{standardregios} ($c$). At nonvanishing $\epsilon $, each cut splits into
two cuts, as shown in fig. \ref{standardregios} ($a$), with branch points $%
p^{0}=\pm (2m^{2}-i\epsilon )/m$ and $p^{0}=\pm (2m^{2}+i\epsilon )/m$. Such
cuts do not intersect the real axis at $\epsilon \neq 0$, so we are allowed
to study the cutting equation in any interval of the real axis and
analytically continue the result to the whole real axis.

The limit $\epsilon \rightarrow 0$ divides the complex plane into
disconnected regions. If we deform the cuts analytically before the limit,
we may obtain fig. \ref{standardregios} ($b$), the disconnected regions
being $\mathcal{A}$ and $\mathcal{A}^{\prime }$. However, what physics
prescribes is fig. \ref{standardregios} ($c$), where the region $\mathcal{A}%
^{\prime }$ is squeezed to the real axis. The value of the amplitude on the
real axis is obtained by approaching the real axis from above, while the
value of the complex conjugate amplitude is obtained by approaching the real
axis from below. Let us inquire about the value of the amplitude in the
region $\mathcal{A}^{\prime }$.

\begin{figure}[tbp]
\begin{center}
\includegraphics[width=16truecm]{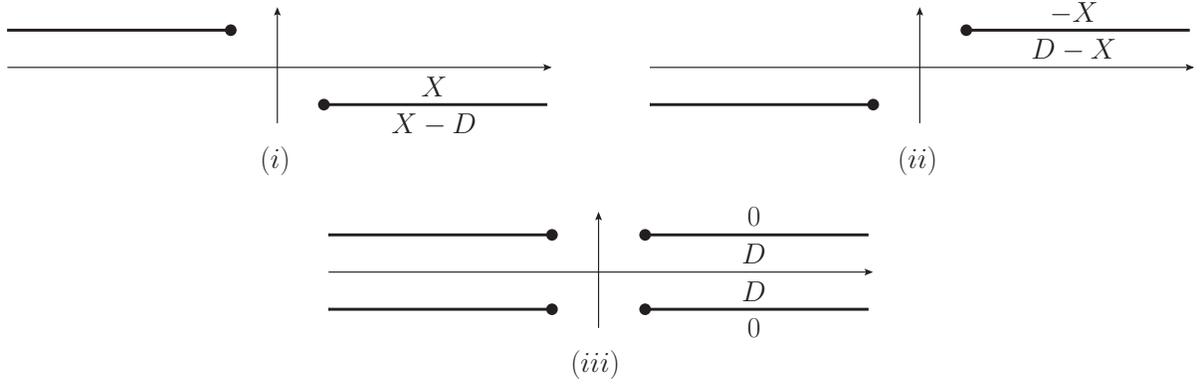}
\end{center}
\caption{Branch cuts of the cutting equations for the standard bubble
diagram }
\label{disco}
\end{figure}

It is easy to show that, in the limit $\epsilon \rightarrow 0$, the value of
the cut diagram in the intersection $\mathcal{A}_{\perp }^{\prime }$ between 
$\mathcal{A}^{\prime }$ and the real axis is equal to the discontinuity of
the amplitude. Indeed, consider fig. \ref{disco}, which shows the complex
energy plane in the case of ($i$) the bubble diagram (fig. \ref{bubble}), ($%
ii$) its conjugate and ($iii$) minus the sum of the two cut diagrams (fig. %
\ref{bubblecut}), respectively. The cuts are displaced from the real axis at 
$\epsilon \neq 0$. In what follows we imagine to take $\epsilon \rightarrow 0
$ and just report the values of the diagrams in this limit. The cutting
equation tells us that the sum $(i)+(ii)$ is equal to ($iii$) and that ($iii$%
) vanishes everywhere except in the cuts for $\epsilon \rightarrow 0$. Let $X
$ denote the value of the bubble diagram above the cut in that limit and $%
D=-1/(8\pi )\sqrt{1-(4m^{2}/p^{2})}$ the well-known value of the
discontinuity. Then, the value of the bubble diagram below the cut is $X-D$.
Since the sum $(i)+(ii)$ must vanish both above the two cuts and below them,
we infer that the value of the conjugate bubble diagram is $-X$ above the
cut and $D-X$ below the cut. This implies that the region in between the
cuts of figure ($iii$) must have value $D$, which is what we claimed. Thus,
the analytic function of the region $\mathcal{A}^{\prime }$ shown in fig. %
\ref{standardregios} ($b$) is $-1/(8\pi )\sqrt{1-(4m^{2}/p^{2})}$, while the
analytic function of the region $\mathcal{A}$ is identically zero.

We see that a single function that is analytic in a neighborhood of the real
axis at $\epsilon \neq 0$ breaks into multiple analytic functions when $%
\epsilon $ tends to zero. Each cut diagram is analytic throughout the real
axis at $\epsilon \neq 0$. Instead, at $\epsilon =0$ the real axis is
divided into several domains and the cut diagram is separately analytic in
each domain.

These remarks are useful when we move to the Lee-Wick models. We consider a
generic one-loop diagram and describe how the cutting equations are derived
in sections \ref{LWbubble} and \ref{LWuntriangle}, focusing on the regions
that have intersections with the real axis. We must combine the discussion
about the analytic regions associated with the LW pinching with the
discussion about the usual pinching. The standard threshold reads $%
p^{2}=(m_{i}+m_{j})^{2}$, where $m_{i}$ and $m_{j}$ are the masses of two
particles circulating in the loop and $p$ is a sum of incoming momenta. The
LW threshold on the real axis is $p^{2}=2M^{2}$.

\begin{figure}[t]
\begin{center}
\includegraphics[width=11truecm]{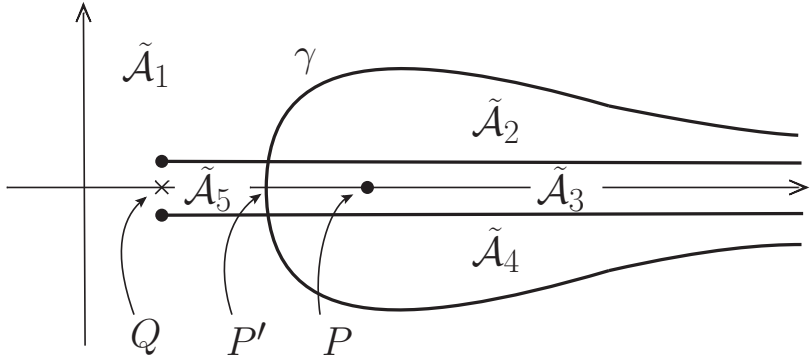}
\end{center}
\caption{Standard pinching and LW pinching}
\label{sigmainter}
\end{figure}

Assume first that $2M^{2}>(m_{i}+m_{j})^{2}$ and $P^{\prime }$ is located
above $Q$. The cut diagram leads to a typical situation like the one of fig. %
\ref{sigmainter}. Above the LW threshold $P$, we study the difference $i%
\mathcal{M}-i\mathcal{M}^{\ast }$, where $\mathcal{M}=-i\lambda ^{2}\mathcal{%
J}/2$ is the amplitude, by working at $\epsilon \neq 0$ in a subdomain of $%
\mathcal{\tilde{A}}_{3}$,\ for example an interval $\mathcal{D}_{3}$ of the
real axis. Then, we perform the domain deformation, which squeezes the
region $\mathcal{\tilde{A}}_{3}$ to the real axis, till it becomes the
portion $\mathcal{A}_{3}=\mathcal{O}_{P}$ of the real axis from $P$ to
infinity. We show that the calculation can be extended through the domain
deformation. Nevertheless, $i\mathcal{M}-i\mathcal{M}^{\ast }$ does not have
the expected form compatible with unitarity, as long as $\epsilon $ remains
nonzero. At the end, we take the limit $\epsilon \rightarrow 0$ and prove
that $i\mathcal{M}-i\mathcal{M}^{\ast }$ can be expressed as predicted by
the unitary cutting equation, encoded in the identity $iT-iT^{\dag
}=-TT^{\dag }$.

Below the threshold $P$, the domain deformation is unnecessary. We split the
calculation in two parts. The limit $\epsilon \rightarrow 0$ makes two
standard poles coincide in $Q$ and divides the positive real axis below $%
P^{\prime }$ into two portions: one portion is the domain $\mathcal{D}_{1}$
that goes from the origin to $Q$, which belongs to the region $\mathcal{%
\tilde{A}}_{1}$; the other portion is the domain $\mathcal{D}_{5}$ that goes
from $Q$ to $P^{\prime }$, which belongs to the region $\mathcal{\tilde{A}}%
_{5}$.

We can prove unitarity in $\mathcal{A}_{1}$ by working in an interval of $%
\mathcal{D}_{1}$, integrating rigidly on the loop space momenta, then taking 
$\epsilon \rightarrow 0$ and analytically continuing the cutting equation to
the whole $\mathcal{D}_{1}$ and then $\mathcal{A}_{1}$.

\begin{figure}[b]
\begin{center}
\includegraphics[width=10truecm]{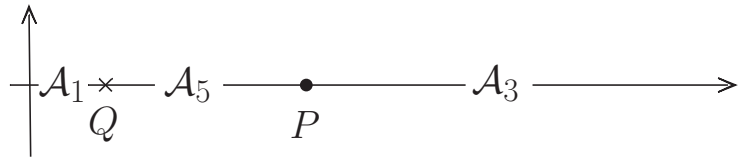}
\end{center}
\caption{Intersections of the analytic regions with the real axis}
\label{sigmainter2}
\end{figure}

Similarly, we can study $i\mathcal{M}-i\mathcal{M}^{\ast }$ in an interval
of $\mathcal{D}_{5}$ (where we are allowed to rigidly integrate on the loop
space momenta, since $\mathcal{D}_{5}$ is below $\gamma $) and then take the
limit $\epsilon \rightarrow 0$. After that, we analytically continue the
cutting equation to the interval of the real axis that goes from $Q$ to $P$.
So doing, we cover the whole region $\mathcal{A}_{5}$, which is the segment
of the real axis going from $Q$ to $P$, including the portion where we
cannot integrate on the loop space momenta rigidly. The analytic regions
that intersect the real axis are then those of fig. \ref{sigmainter2}.

Now, assume that $2M^{2}>(m_{i}+m_{j})^{2}$, but $P^{\prime }$ is located
below $Q$. In that case, it seems that we cannot treat the portion of the
real axis included from $Q$ to $P$ with the method explained above.
Nevertheless, it is always possible to switch to a situation like the one of
fig. \ref{sigmainter}. Indeed, for $\mathbf{p}=0$, $P^{\prime }$ coincides
with $P$ and $Q$ is below $P^{\prime }=P$. By continuity, for nonvanishing,
but sufficiently small $\mathbf{p}$, $Q$ is still below $P^{\prime }$. This
proves that there exists an open domain $\mathcal{O}_{5}\subset \mathcal{%
\tilde{A}}_{5}$ of the space of the external momenta where the point $Q$ is
located below $P^{\prime }$. From $\mathcal{O}_{5}$ we can proceed as
explained above and reach $\mathcal{A}_{5}$ after the analytic continuation.

Finally, when $2M^{2}<(m_{i}+m_{j})^{2}$ the point $Q$ is located above $P$.
Then, below $Q$ we can proceed as in the pure LW\ case, while above $Q$ we
can perform the domain deformation and let $\epsilon $ tend to zero at the
end.

\section{The standard bubble diagram revisited}

\setcounter{equation}{0}

\label{LWun}

In this section, we reconsider the standard bubble diagram and study its
discontinuity. We generalize the usual derivation \cite{peskin} in various
directions, to prepare the extension to the Lee-Wick models.

We use the dimensional regularization and work in a generic Lorentz frame,
instead of choosing, say, the external momentum $p=(p^{0},\mathbf{p})$ of
the form $(p^{0},0)$. One reason is that this choice is only allowed for
timelike external momenta. More importantly, we have seen that in the LW
models it is crucial to keep the external space momentum $\mathbf{p}$
different from zero, to enlarge the region $\mathcal{\tilde{A}}_{P}$ of the
complex plane, which is otherwise squeezed on the real axis.

We also take different masses $m_{1}$, $m_{2}$, and independent
infinitesimal widths $\epsilon _{1}$, $\epsilon _{2}$, which we keep
nonvanishing as long as we can. As shown in ref. \cite{ACE}, it is possible
to work out more general versions of the cutting equations at $\epsilon \neq
0$ in the standard case. We will see in the next sections that this is no
longer true in the LW case.

The loop integral reads 
\begin{equation}
i\mathcal{M}(p)=\frac{\lambda ^{2}}{2}\int \frac{\mathrm{d}^{D}k}{(2\pi )^{D}%
}\frac{1}{k^{2}-m_{1}^{2}+i\epsilon _{1}}\frac{1}{(k-p)^{2}-m_{2}^{2}+i%
\epsilon _{2}},  \label{oneloopstandard}
\end{equation}%
where $\mathcal{M}(p)$ is the amplitude. We can equivalently write (\ref%
{oneloopstandard}) as 
\begin{equation}
i\mathcal{M}(p)=\frac{\lambda ^{2}}{2}\int \frac{\mathrm{d}k^{0}\mathrm{d}%
^{D-1}\mathbf{k}}{(2\pi )^{D}}\prod\limits_{j=1}^{2}\frac{1}{(e_{j}-\omega
_{j}+i\epsilon _{j})(e_{j}+\omega _{j}-i\epsilon _{j})},  \label{olb}
\end{equation}%
where $e_{1}=k^{0}$, $e_{2}=k^{0}-p^{0}$, $\omega _{1}=\sqrt{\mathbf{k}%
{}^{2}+m_{1}^{2}}$ and $\omega _{2}=\sqrt{(\mathbf{k}-\mathbf{p}%
){}^{2}+m_{2}^{2}}$. In going from (\ref{oneloopstandard}) to (\ref{olb}),
we have expanded the\ denominators for $\epsilon _{1}$, $\epsilon _{2}$
small and rescaled such widths.

We perform the integral on $k^{0}$ by using the residue theorem and closing
the integration path in the lower half $k^{0}$ plane. The relevant poles are
located at $k^{0}=z_{1}$ and $k^{0}=z_{2}$, where 
\begin{equation*}
z_{1}=\omega _{1}-i\epsilon _{1},\qquad z_{2}=p^{0}+\omega _{2}-i\epsilon
_{2}.
\end{equation*}%
The $k^{0}$ integral of $i\mathcal{M}$ leads to%
\begin{equation}
i\mathcal{M}(p)=-\frac{i\lambda ^{2}}{2}\int \frac{\mathrm{d}^{D-1}\mathbf{k}%
}{(2\pi )^{D-1}}\left[ \mathrm{Res}(z_{1})+\mathrm{Res}(z_{2})\right] ,
\label{ims}
\end{equation}%
where \textrm{Res}$(z)$ denotes the residue of the integrand (excluding the
factor $\lambda ^{2}/2$) in $k^{0}=z$. We find%
\begin{eqnarray}
\mathrm{Res}(z_{1}) &=&\frac{1}{2z_{1}(z_{1}-z_{2})(z_{1}+z_{2}-2p^{0})}=%
\frac{1}{2(\omega _{1}-i\epsilon _{1})\Delta _{-}(\omega _{1}+\omega
_{2}-p^{0}-i\epsilon _{+})}\equiv r_{1},  \notag \\
\mathrm{Res}(z_{2}) &=&-\frac{1}{2(z_{2}-p^{0})(z_{1}-z_{2})(z_{1}+z_{2})}=-%
\frac{1}{2(\omega _{2}-i\epsilon _{2})\Delta _{-}(\omega _{1}+\omega
_{2}+p^{0}-i\epsilon _{+})}\equiv r_{2},\qquad   \label{resi}
\end{eqnarray}%
while 
\begin{equation}
\Delta _{-}=z_{1}-z_{2}=\omega _{1}-\omega _{2}-p^{0}-i\epsilon _{-}
\label{deltam}
\end{equation}%
and $\epsilon _{\pm }=\epsilon _{1}\pm \epsilon _{2}$. The denominator $%
\Delta _{-}$ gives ambiguous distributions, since the sign of $\epsilon _{-}$
depends on the order with which we perform the limits $\epsilon
_{1}\rightarrow 0$ and $\epsilon _{2}\rightarrow 0$. As shown in ref. \cite%
{ACE}, the ambiguity must actually cancel out. Indeed, it does disappear as
soon as we take the sum of the two residues, which gives%
\begin{eqnarray}
\mathrm{Res}(z_{1})+\mathrm{Res}(z_{2}) &=&-\frac{1}{4z_{1}(z_{2}-p^{0})}%
\left( \frac{1}{z_{1}+z_{2}-2p^{0}}+\frac{1}{z_{1}+z_{2}}\right)   \notag \\
&=&-\frac{1}{4\omega _{1}\omega _{2}}\left( \frac{1}{\omega _{1}+\omega
_{2}-p^{0}-i\epsilon _{+}}+\frac{1}{\omega _{1}+\omega _{2}+p^{0}-i\epsilon
_{+}}\right) .  \label{rr}
\end{eqnarray}%
In the last line we sent $\epsilon _{1}$ and $\epsilon _{2}$ to zero in a
couple of places where they are unimportant. For example, the factor $1/z_{1}
$ can be replaced by $1/\omega _{1}$. It is\ not convenient to make this
replacement directly in formulas (\ref{resi}), because of the presence of
the ambiguous denominator (\ref{deltam}). In the rest of the paper, we make
similar replacements, when they are allowed, without further notice.

The discontinuity $\mathrm{Disc}\mathcal{M}=2i\mathrm{Im}\mathcal{M}$ of the
amplitude can now be evaluated from (\ref{ims}) by means of the identity 
\begin{equation}
\frac{1}{x\pm i\epsilon }=\mathcal{P}\left( \frac{1}{x}\right) \mp i\pi
\delta (x),  \label{deco}
\end{equation}%
where\ $\mathcal{P}$ denotes the principal value. We find $\mathrm{Disc}%
\mathcal{M}=i\lambda ^{2}\Upsilon /2$, where 
\begin{equation}
\Upsilon (p)\equiv \int \frac{\mathrm{d}^{D-1}\mathbf{k}}{(2\pi )^{D-1}}%
\frac{2\pi }{(2\omega _{1})(2\omega _{2})}\left[ \delta (p^{0}-\omega
_{1}-\omega _{2})+\delta (p^{0}+\omega _{1}+\omega _{2})\right] .
\label{integral}
\end{equation}

\begin{figure}[t]
\begin{center}
\includegraphics[width=9truecm]{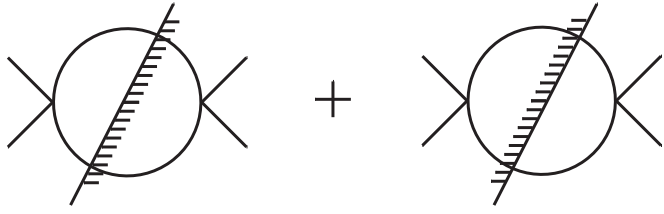}
\end{center}
\caption{Cut bubble diagrams}
\label{bubblecut}
\end{figure}

If we relabel $q_{1}=k$ and $q_{2}=p-k$ and introduce integrals over $q_{1}$
and $q_{2}$, together with delta functions that impose $q_{i}^{0}=\pm \omega
_{i}^{\prime }$, where $\omega _{i}^{\prime }\equiv \sqrt{\mathbf{q}%
_{i}^{2}+m_{i}^{2}}$, and $\mathbf{p}=\mathbf{q}_{1}+\mathbf{q}_{2}$, we can
view $\delta (p^{0}\pm \omega _{1}\pm \omega _{2})$ as the total energy
conservation $\delta (p^{0}-q_{1}^{0}-q_{2}^{0})$. Then we can write $%
\Upsilon $ as 
\begin{equation*}
\int \frac{\mathrm{d}^{D}q_{1}}{(2\pi )^{D}}\frac{\mathrm{d}^{D}q_{2}}{(2\pi
)^{D}}(2\pi )^{D}\delta ^{(D)}(p-q_{1}-q_{2})\left[ \frac{2\pi \delta
(q_{1}^{0}-\omega _{1}^{\prime })2\pi \delta (q_{2}^{0}-\omega _{2}^{\prime
})}{(2\omega _{1}^{\prime })(2\omega _{2}^{\prime })}+\frac{2\pi \delta
(q_{1}^{0}+\omega _{1}^{\prime })2\pi \delta (q_{2}^{0}+\omega _{2}^{\prime
})}{(2\omega _{1}^{\prime })(2\omega _{2}^{\prime })}\right]
\end{equation*}%
and finally 
\begin{equation}
\int \frac{\mathrm{d}^{D}q_{1}}{(2\pi )^{D}}\frac{\mathrm{d}^{D}q_{2}}{(2\pi
)^{D}}(2\pi )^{D}\delta ^{(D)}(p-q_{1}-q_{2})(2\pi )\delta
(q_{1}^{2}-m_{1}^{2})(2\pi )\delta (q_{2}^{2}-m_{2}^{2})\left[ \theta
(q_{1}^{0})\theta (q_{2}^{0})+\theta (-q_{1}^{0})\theta (-q_{2}^{0})\right] .
\label{final}
\end{equation}%
We see that $\mathrm{Disc}\mathcal{M}=i\lambda ^{2}\Upsilon /2$ is equal to $%
i$ times the sum of the two cut diagrams $\mathcal{C}_{1}$, $\mathcal{C}_{2}$
shown in fig. \ref{bubblecut}, i.e. 
\begin{equation}
i\mathcal{M}-i\mathcal{M}^{\ast }=-\frac{\lambda ^{2}}{2}\Upsilon =-\mathcal{%
C}_{1}-\mathcal{C}_{2}.  \label{bubblecutf}
\end{equation}%
The cut diagrams can be computed by replacing the ordinary propagators with
the cut ones, 
\begin{equation}
\frac{i}{p^{2}-m^{2}+i\epsilon }\rightarrow 2\pi \theta (\pm p^{0})\delta
(p^{2}-m^{2}),  \label{complcut}
\end{equation}%
and equipping each shadowed vertex with a minus sign. The sign in front of $%
p^{0}$ is determined by the direction of the energy flow through the cut.

Formula (\ref{bubblecutf}) is nothing but the relation $iT-iT^{\dag
}=-TT^{\dag }$ in the particular case we are considering and shows that the
bubble diagram satisfies unitarity.

For completeness, we report the value of the integral $\Upsilon $ in four
dimensions, which is well known:%
\begin{equation}
\Upsilon (p)=\frac{1}{8\pi p^{2}}\theta (p^{2}-(m_{1}+m_{2})^{2})\sqrt{%
\left( p^{2}-m_{1}^{2}-m_{2}^{2}\right) ^{2}-4m_{1}^{2}m_{2}^{2}}.
\label{Ires}
\end{equation}

\subsection{A. Comments}

The procedure we have used in this section is general enough to be extended\
to the LW models. However, before moving to the LW case, we would like to
emphasize the strategy of the calculation and compare it with other
strategies.

The first step has been to integrate on the energy by means of the residue
theorem. \textit{Only after that}, we have used the decomposition (\ref{deco}%
). The usage of that decomposition is extremely delicate, especially in
products of distributions. For example, it is very inconvenient to use it
before applying the residue theorem, directly in formula (\ref%
{oneloopstandard}). If we do so (working at $\mathbf{p}=0$ and in the equal
mass case $m_{1}=m_{2}=m$, for simplicity), we get 
\begin{equation}
\mathrm{Disc}\mathcal{M}=i\lambda ^{2}\int \frac{\mathrm{d}^{D}k}{(2\pi )^{D}%
}\left[ \pi ^{2}\delta (k^{2}-m^{2})\delta ((p-k)^{2}-m^{2})-\mathcal{P}%
\frac{1}{k^{2}-m^{2}}\mathcal{P}\frac{1}{(k-p)^{2}-m^{2}}\right] .
\label{wro}
\end{equation}%
The first contribution to (\ref{wro}), which can be rewritten as 
\begin{equation}
\frac{i\lambda ^{2}}{4}\int \frac{\mathrm{d}^{D}q_{1}}{(2\pi )^{D}}\frac{%
\mathrm{d}^{D}q_{2}}{(2\pi )^{D}}(2\pi )^{D}\delta
^{(D)}(p-q_{1}-q_{2})(2\pi )\delta (q_{1}^{2}-m^{2})(2\pi )\delta
(q_{2}^{2}-m^{2}),  \label{wrong}
\end{equation}%
resembles the final result $i\lambda ^{2}\Upsilon /2$, with $\Upsilon $
given by (\ref{final}). This lead some authors \cite{shapiromodesto} to
think that the two are equal and that the term with the two principal values
in (\ref{wro}) does not contribute. Both statements are incorrect.

An apparent difference between $i\lambda ^{2}\Upsilon /2$ and (\ref{wrong})
is that (\ref{final}) contains the combination of theta functions $\Theta
\equiv \theta (q_{1}^{0})\theta (q_{2}^{0})+\theta (-q_{1}^{0})\theta
(-q_{2}^{0})$, while (\ref{wrong}) does not. If we multiply the integrand of
(\ref{wrong}) by $1=\Theta +1-\Theta $, we can easily check that the
difference $1-\Theta $, which is equal to $\theta (q_{1}^{0})\theta
(-q_{2}^{0})+\theta (q_{1}^{0})\theta (-q_{2}^{0})$, gives zero. Thus, we
can safely insert $\Theta $ in the integral of (\ref{wrong}) and make it
more similar to $i\lambda ^{2}\Upsilon /2$.

A more serious difference, instead, is the multiplying factor. Formula (\ref%
{wrong}) is not really equal to $i\lambda ^{2}/2$ times (\ref{final}),
contrary to the claim of ref. \cite{shapiromodesto}, because it is
multiplied by an additional factor $1/2$. The missing contribution must come
from the product of the two principal values in formula (\ref{wro}). We have
checked this fact numerically with a \textit{Mathematica} program, starting
from 
\begin{equation*}
-i\lambda ^{2}\int \frac{\mathrm{d}^{D}k}{(2\pi )^{D}}\frac{k^{2}-m^{2}}{%
(k^{2}-m^{2})^{2}+\epsilon ^{2}}\frac{(p-k)^{2}-m^{2}}{%
((p-k)^{2}-m^{2})^{2}+\epsilon ^{2}}
\end{equation*}%
and taking smaller and smaller values of $\epsilon $. The argument used in
ref. \cite{shapiromodesto} to claim that this expression vanishes was to
turn it to the Euclidean framework, where it naively becomes real, while in
Minkowski spacetime it is purely imaginary. The point is that the Wick
rotation is nontrivial in this case, because the integrand has poles in the
first and third quadrants, which must be taken into account. More details
can be found in ref. \cite{ugo}, where the problems of these types of
Minkowski integrals are studied in depth.

\section{The Lee-Wick bubble diagram}

\label{LWbubble}

In this section we study the LW version of the bubble diagram and show that
it satisfies the correct cutting equation, with no propagation of unphysical
degrees of freedom through the cuts. The loop integral is 
\begin{eqnarray}
i\mathcal{M} &=&\frac{\lambda ^{2}}{2}\int \frac{\mathrm{d}^{D}k}{(2\pi )^{D}%
}D(k^{2},m_{1}^{2},\epsilon _{1})D((k-p)^{2},m_{2}^{2},\epsilon _{2})
\label{bubblo} \\
&=&\frac{\lambda ^{2}M^{8}}{2}\int \frac{\mathrm{d}k^{0}\mathrm{d}^{D-1}%
\mathbf{k}}{(2\pi )^{D}}\prod\limits_{j=1}^{2}\frac{1}{(e_{j}-\nu
_{j})(e_{j}+\nu _{j})(e_{j}-\nu _{j}^{\ast })(e_{j}+\nu _{j}^{\ast
})(e_{j}-\omega _{j}+i\epsilon _{j})(e_{j}+\omega _{j}-i\epsilon _{j})}, 
\notag
\end{eqnarray}%
where $\nu _{1}=\sqrt{\mathbf{k}^{2}+iM^{2}}$, $\nu _{2}=\sqrt{(\mathbf{k}-%
\mathbf{p})^{2}+iM^{2}}$ and the other definitions coincide with those of
the previous section. For the reasons explained in section \ref{LWE}, it is
important to work at $\mathbf{p}\neq 0$.

Making the Wick rotation and closing the $k^{0}$ integration path in the
lower half plane, fig. \ref{WickBub2} tells us that we need the poles%
\begin{equation}
z_{1}=\omega _{1}-i\epsilon _{1},\qquad z_{2}=p^{0}+\omega _{2}-i\epsilon
_{2},\qquad w_{1}=\nu _{1},\qquad w_{2}=p^{0}+\nu _{2},  \label{reso}
\end{equation}%
together with the conjugates $w_{1}^{\ast }$ and $w_{2}^{\ast }$. We find 
\begin{equation}
i\mathcal{M}=-\frac{i\lambda ^{2}}{2}\int \frac{\mathrm{d}^{D-1}\mathbf{k}}{%
(2\pi )^{D-1}}[\mathrm{Res}(z_{1})+\mathrm{Res}(z_{2})+\mathrm{Res}(w_{1})+%
\mathrm{Res}(w_{2})+\mathrm{Res}(w_{1}^{\ast })+\mathrm{Res}(w_{2}^{\ast })].
\label{imlw}
\end{equation}

We perform the domain deformations associated with the contributions of $%
\mathrm{Res}(w_{i})$ and $\mathrm{Res}(w_{i}^{\ast })$ in complex conjugate
ways. Then, calling $U$ and $U^{\ast }$ the deformed domains, such
contributions read%
\begin{equation}
-\frac{i\lambda ^{2}}{2}\int_{U}\frac{\mathrm{d}^{D-1}\mathbf{k}}{(2\pi
)^{D-1}}\mathrm{Res}(w_{i})-\frac{i\lambda ^{2}}{2}\int_{U^{\ast }}\frac{%
\mathrm{d}^{D-1}\mathbf{k}}{(2\pi )^{D-1}}\mathrm{Res}(w_{i}^{\ast }).
\label{dd}
\end{equation}%
The other contributions, due to $\mathrm{Res}(z_{i})$, can be calculated
with the natural real $\mathbf{k}$ integration domain.

Now, we can write%
\begin{equation}
\left[ \int_{U^{\ast }}\frac{\mathrm{d}^{D-1}\mathbf{k}}{(2\pi )^{D-1}}%
\mathrm{Res}(w_{i}^{\ast })\right] ^{\ast }=\int_{U}\frac{\mathrm{d}^{D-1}%
\mathbf{k}}{(2\pi )^{D-1}}[\mathrm{Res}(w_{i}^{\ast })]^{\ast },  \label{hy}
\end{equation}%
where it is understood that the complex conjugations in $[\mathrm{Res}%
(w_{i}^{\ast })]^{\ast }$ do not act on $\mathbf{k}$. We prove the identity 
\begin{equation}
\lbrack \mathrm{Res}(w_{i}^{\ast })]^{\ast }=\mathrm{Res}(w_{i})
\label{crucial}
\end{equation}%
at $\epsilon _{1}=\epsilon _{2}=0$, which allows us to turn (\ref{hy}) into%
\begin{equation}
\left[ \int_{U^{\ast }}\frac{\mathrm{d}^{D-1}\mathbf{k}}{(2\pi )^{D-1}}%
\mathrm{Res}(w_{i}^{\ast })\right] ^{\ast }=\int_{U}\frac{\mathrm{d}^{D-1}%
\mathbf{k}}{(2\pi )^{D-1}}\mathrm{Res}(w_{i}).  \label{compu}
\end{equation}

Formula (\ref{crucial}) expresses the compensation between the contributions
of the poles of the same LW pair to the cutting equations. It is the key
result to prove that only the physical degrees of freedom propagate through
the cuts.

To derive (\ref{crucial}), observe that when $\epsilon _{1}$ and $\epsilon
_{2}$ tend to zero we have%
\begin{equation}
\mathrm{Res}(w_{1})-[\mathrm{Res}(w_{1}^{\ast })]^{\ast }=\frac{\pi M^{6}}{%
2(m_{1}^{2}-iM^{2})\nu _{1}}\frac{\tilde{\delta}((\nu _{1}-p^{0})^{2}-\omega
_{2}^{2})}{((\nu _{1}-p^{0})^{2}-(\mathbf{k}-\mathbf{p})^{2})^{2}+M^{4}},
\label{ress}
\end{equation}%
where%
\begin{equation*}
\tilde{\delta}(z)\equiv \lim_{\epsilon \rightarrow 0}\Delta _{\epsilon
}(z),\qquad \Delta _{\epsilon }(z)\equiv \frac{1}{2i\pi }\left( \frac{1}{%
z-i\epsilon }-\frac{1}{z+i\epsilon }\right) ,
\end{equation*}%
is the usual delta distribution extended to complex values, which means that
it is equal to zero anywhere but on the real axis, where it is the ordinary
delta function. The right-hand side of (\ref{ress}) collects the terms where
the limits $\epsilon _{1},\epsilon _{2}\rightarrow 0$ are nontrivial, which
need to be studied in detail.

Now we show that $\tilde{\delta}(z)$ does not contribute to the integrals (%
\ref{dd}). The pinching condition for both $\mathrm{Res}(w_{1})$ and $[%
\mathrm{Res}(w_{1}^{\ast })]^{\ast }$ is (\ref{pinchcond}), i.e. $p^{0}=\nu
_{1}+\nu _{2}^{\ast }$, where, again, the complex conjugation does not act
on $\mathbf{k}$. On the other hand, the argument of $\tilde{\delta}$
vanishes for $p^{0}=\nu _{1}\pm \omega _{2}$. These conditions cannot hold
at the same time, because $p^{0}=\nu _{1}+\nu _{2}^{\ast }=\nu _{1}\pm
\omega _{2}$ implies $-iM^{2}=m_{2}^{2}$. This fact has important
consequences. When $\epsilon $ tends to zero, the contributions to $\Delta
_{\epsilon }(z)$ provide potential singularities $\sim 1/z$, with $z=(\nu
_{1}-p^{0})^{2}-\omega _{2}^{2}$. However, such singularities are actually
integrable, because $z$ is complex. Therefore, we have two potential
singularities, those due to the LW\ pinching and those due to $1/z$. Both
are separately integrable and could only give trouble if they occurred at
the same time. Since this is impossible, the two contributions to $\Delta
_{\epsilon }(z)$ mutually cancel for $\epsilon \rightarrow 0$ and the
right-hand side of (\ref{ress}) can be dropped.

Similar arguments can be applied to $\mathrm{Res}\mathbb{(}w_{2})-[\mathrm{%
Res}(w_{2}^{\ast })]^{\ast }$. We conclude that identity (\ref{crucial})
holds. Thanks to it, the second integral of (\ref{dd}) is the complex
conjugate of the first integral, so (\ref{compu}) holds.

Since formula (\ref{crucial}) is valid only at $\epsilon _{1}=\epsilon _{2}=0
$, we have to explain when such widths must be sent to zero. We work in the
intervals $\mathcal{D}_{1}$, $\mathcal{D}_{3}$ and $\mathcal{D}_{5}$ defined
in section \ref{anacut}. In $\mathcal{D}_{3}$ we perform the domain
deformation at $\epsilon _{1},\epsilon _{2}\neq 0$, for the contributions
due to $\mathrm{Res}(w_{i})$ and $\mathrm{Res}(w_{i}^{\ast })$. Instead, we
keep the $\mathbf{k}$ integration domain rigid for the other contributions,
as well as for the calculations in $\mathcal{D}_{1}$ and $\mathcal{D}_{5}$.
Then, by means of identities like (\ref{ress}) and the calculations reported
below, we check that the expected cutting equation separately holds in $%
\mathcal{D}_{1}$, $\mathcal{D}_{3}$ and $\mathcal{D}_{5}$, up to corrections
of the form $\Delta _{\epsilon }(z)$, which are killed by the limit $%
\epsilon \rightarrow 0$. From section \ref{anacut}, we know that the cutting
equation can be analytically extended from $\mathcal{D}_{1}$, $\mathcal{D}%
_{3}$ and $\mathcal{D}_{5}$ to the regions $\mathcal{A}_{1}$, $\mathcal{A}%
_{3}$ and $\mathcal{A}_{5}$, respectively.

Taking the limit $\epsilon _{1},\epsilon _{2}\rightarrow 0$ on the
contributions of $\mathrm{Res}(w_{i})$ and $\mathrm{Res}(w_{i}^{\ast })$ to (%
\ref{imlw}), but keeping $\epsilon _{1},\epsilon _{2}\neq 0$ in the
contributions due to $\mathrm{Res}(z_{1})$ and $\mathrm{Res}(z_{2})$,
formulas (\ref{dd}) and (\ref{compu})\ give%
\begin{equation*}
i\mathcal{M}=-\frac{i\lambda ^{2}}{2}\int \frac{\mathrm{d}^{D-1}\mathbf{k}}{%
(2\pi )^{D-1}}\left[ \mathrm{Res}(z_{1})+\mathrm{Res}(z_{2})\right]
-i\lambda ^{2}\mathrm{Re}\int_{U}\frac{\mathrm{d}^{D-1}\mathbf{k}}{(2\pi
)^{D-1}}[\mathrm{Res}(w_{1})+\mathrm{Res}(w_{2})].
\end{equation*}%
The discontinuity of the amplitude is then%
\begin{equation*}
\mathrm{Disc}\mathcal{M}=2i\mathrm{Im}\mathcal{M=}-i\lambda ^{2}\int \frac{%
\mathrm{d}^{D-1}\mathbf{k}}{(2\pi )^{D-1}}\mathrm{Im}\left[ \mathrm{Res}%
(z_{1})+\mathrm{Res}(z_{2})\right] .
\end{equation*}%
This result proves that, as anticipated, the LW poles do not contribute to
the imaginary part of the amplitude.

Now we show that the amplitude obeys the correct cutting equation. We have

\begin{equation}
\text{\textrm{Res}}(z_{1})=r_{1}h(z_{1}),\qquad \mathrm{Res}%
(z_{2})=r_{2}h(z_{2}),  \label{resa}
\end{equation}%
where $r_{1}$ and $r_{2}$ are defined in formula (\ref{resi}) and%
\begin{equation*}
h(k^{0})=\frac{M^{4}}{(k^{2})^{2}+M^{4}}\frac{M^{4}}{((k-p)^{2})^{2}+M^{4}}.
\end{equation*}%
We understand the dependence of $h$ on the other variables besides $k^{0}$,
because they are not important for the discussion.

As before, the ill-defined distributions contained in $r_{1}$ and $r_{2}$
cancel out in the sum of \textrm{Res}$(z_{1})$ and $\mathrm{Res}(z_{2})$. We
have

\begin{equation}
\text{\textrm{Res}}(z_{1})+\mathrm{Res}%
(z_{2})=r_{1}h(z_{1})+r_{2}h(z_{2})=u(z_{1},z_{2})v(z_{1},z_{2}),
\label{total}
\end{equation}%
where%
\begin{eqnarray*}
u(z_{1},z_{2}) &=&\frac{%
z_{1}(z_{2}-p^{0})[h(z_{1})-h(z_{2})]+z_{2}(z_{2}-p^{0})h(z_{1})-z_{1}(z_{1}-p^{0})h(z_{2})%
}{4p^{0}(z_{1}-z_{2})z_{1}(z_{2}-p^{0})}, \\
v(z_{1},z_{2}) &=&\frac{1}{z_{1}+z_{2}-2p^{0}}-\frac{1}{z_{1}+z_{2}}.
\end{eqnarray*}%
It is clear that $u(z_{1},z_{2})$ is regular, since the numerator vanishes
when $z_{1}=z_{2}$. Note that $h(z)$ is real and nonvanishing for real $z$.
Thus, we can replace $u(z_{1},z_{2})$ by $u(\omega _{1},\omega _{2}+p^{0})$.
At this point, we just need to take the imaginary part of $v(z_{1},z_{2})$
by means of formula (\ref{deco}), which gives%
\begin{eqnarray}
\mathrm{Disc}\mathcal{M} &=&-i\pi \lambda ^{2}\int \frac{\mathrm{d}^{D-1}%
\mathbf{k}}{(2\pi )^{D-1}}u(\omega _{1},\omega _{2}+p^{0})\left[ \delta
(\omega _{1}+\omega _{2}-p^{0})-\delta (\omega _{1}+\omega _{2}+p^{0})\right]
\notag \\
&=&\frac{i\lambda ^{2}}{2}\frac{\,M^{8}}{(M^{4}+m_{1}^{4})(M^{4}+m_{2}^{4})}%
\Upsilon ,  \label{discM}
\end{eqnarray}%
where $\Upsilon $ is the integral (\ref{integral}). The second line is
obtained by noting that the delta functions that appear in the first line of
(\ref{discM}) simplify the value of the function $u$ considerably and allow
us to make the replacements%
\begin{equation*}
u(\omega _{1},\omega _{2}+p^{0})\delta (\omega _{1}+\omega _{2}\pm
p^{0})\rightarrow \pm \frac{1}{4\omega _{1}\omega _{2}}\frac{\,M^{8}\delta
(\omega _{1}+\omega _{2}\pm p^{0})}{(M^{4}+m_{1}^{4})(M^{4}+m_{2}^{4})}.
\end{equation*}

Following the procedure that we used in the standard case, we relabel $%
q_{1}=k$ and $q_{2}=p-k$ and obtain that $\mathrm{Disc}\mathcal{M}$ is equal
to $i\lambda ^{2}/2$ times%
\begin{equation*}
\int \frac{\mathrm{d}^{D}q_{1}}{(2\pi )^{D}}\frac{\mathrm{d}^{D}q_{2}}{(2\pi
)^{D}}(2\pi )^{D}\delta
^{(D)}(p-q_{1}-q_{2})D_{c}(q_{1}^{2},m_{1}^{2})D_{c}(q_{2}^{2},m_{2}^{2})%
\left[ \theta (q_{1}^{0})\theta (q_{2}^{0})+\theta (-q_{1}^{0})\theta
(-q_{2}^{0})\right] ,
\end{equation*}%
where 
\begin{equation}
D_{c}(p^{2},m^{2})=2\pi \delta (p^{2}-m^{2})\frac{M^{4}}{M^{4}+m^{4}}.
\label{dc}
\end{equation}

So doing, we have shown that (\ref{bubblecutf}) holds in each interval $%
\mathcal{D}_{1}$, $\mathcal{D}_{3}$ and $\mathcal{D}_{5}$ of the real axis,
with the cut propagators $\theta (\pm p^{0})D_{c}(p^{2},m^{2})$. Then, we
analytically continue (\ref{bubblecutf}) to the regions $\mathcal{A}_{1}$, $%
\mathcal{A}_{3}$ and $\mathcal{A}_{5}$. Unitarity is verified, because the
cut propagators $\theta (\pm p^{0})D_{c}(p^{2},m^{2})$ just propagate the
physical degrees of freedom.

\section{The LW\ triangle diagram}

\setcounter{equation}{0}\label{LWuntriangle}

\begin{figure}[b]
\begin{center}
\includegraphics[width=2.8truecm]{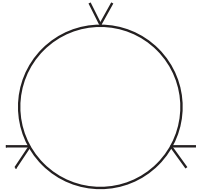}
\end{center}
\caption{Triangle diagram}
\label{triangled}
\end{figure}

In this section we prove that the triangle diagram (fig. \ref{triangled})
also satisfies the correct cutting equation. The loop integral reads 
\begin{eqnarray*}
i\mathcal{M} &=&\lambda ^{3}\int \frac{\mathrm{d}^{D}k}{(2\pi )^{D}}%
D(k^{2},m_{1}^{2},\epsilon _{1})D((k-p)^{2},m_{2}^{2},\epsilon
_{2})D((k-q)^{2},m_{3}^{2},\epsilon _{3}) \\
&=&\lambda ^{3}M^{12}\int \frac{\mathrm{d}k^{0}\mathrm{d}^{D-1}\mathbf{k}}{%
(2\pi )^{D}}\prod\limits_{j=1}^{3}\frac{1}{(e_{j}-\nu _{j})(e_{j}+\nu
_{j})(e_{j}-\nu _{j}^{\ast })(e_{j}+\nu _{j}^{\ast })(e_{j}-\omega
_{j}+i\epsilon _{j})(e_{j}+\omega _{j}-i\epsilon _{j})},
\end{eqnarray*}%
where $e_{3}=k^{0}-q^{0}$, $\omega {}_{3}=\sqrt{(\mathbf{k-q})^{2}+m_{3}^{2}}
$, $\nu _{3}=\sqrt{(\mathbf{k-q})^{2}+iM^{2}}$ and the other definitions
coincide with those of the sections \ref{LWun} and \ref{LWbubble}.

Closing the $q^{0}$ integration path in the lower half plane, we just need
the residues of the poles 
\begin{equation*}
z_{1}=\omega _{1}-i\epsilon _{1},\qquad z_{2}=p^{0}+\omega _{2}-i\epsilon
_{2},\qquad z_{3}=q^{0}+\omega _{3}-i\epsilon _{3},
\end{equation*}%
\begin{equation*}
w_{1}=\nu _{1},\qquad w_{2}=p^{0}+\nu _{2},\qquad w_{3}=q^{0}+\nu _{3},
\end{equation*}%
together with $w_{1}^{\ast }$, $w_{2}^{\ast }$ and $w_{3}^{\ast }$.

The LW thresholds that are located on the real axis are 
\begin{equation*}
p^{2}=2M^{2},\qquad q^{2}=2M^{2},\qquad (p-q)^{2}=2M^{2}.
\end{equation*}%
As explained in section \ref{LWE}, we must work at $\epsilon \neq 0$,
choosing generic external momenta $p$ and $q$ in a generic Lorentz frame. In
each region $\mathcal{\tilde{A}}_{i}$ of the space of $p$ and $q$, we choose
one or more subdomains $\mathcal{O}_{i}$ with an accumulation point,
typically intervals of the real axis. We separate the contributions where it
is necessary to deform the integration domain of the loop space momentum $%
\mathbf{k}$ from the contributions where the deformation is unnecessary. It
can be easily checked that an identity of the form (\ref{ress}) still holds.
Like before, the right-hand side of (\ref{ress}) can be dropped, because the
potential singularities $\sim 1/z$ are integrable and do not occur
simultaneously with the LW\ pinching. This leads again to the crucial
cancellation formula (\ref{crucial}), which ensures that only the standard
residues contribute to the imaginary part of $\mathcal{M}$, in the limit $%
\epsilon \rightarrow 0$. At the end, we analytically continue the cutting
equation from the subdomains $\mathcal{O}_{i}$ to the whole analytic regions 
$\mathcal{A}_{i}$.

Some attention must be paid to the ill-defined distributions, which are more
tricky than in the previous case.

\subsection*{A. The ill-defined distributions cancel out again}

The method we use here to prove this result is simpler than the one of the
previous section, but we have to take the limit $\epsilon \rightarrow 0$ at
a slightly earlier stage.

The residue of the integrand (excluding the factor $\lambda ^{3}$) in $z_{1}$
is%
\begin{eqnarray}
&&\frac{M^{12}}{2\omega _{1}(m_{1}^{4}+M^{4})}\frac{1}{|(\omega
_{1}-p^{0})^{2}-\nu _{2}^{2}|^{2}|(\omega _{1}-q^{0})^{2}-\nu _{3}^{2}|^{2}}%
\frac{1}{(\omega _{1}+\omega _{2}-p^{0}-i\epsilon _{+}^{12})(\omega
_{1}+\omega _{3}-q^{0}-i\epsilon _{+}^{13})}  \notag \\
&&\qquad \qquad \qquad \qquad \times \frac{1}{(\omega _{1}-\omega
_{2}-p^{0}-i\epsilon _{-}^{12})(\omega _{1}-\omega _{3}-q^{0}-i\epsilon
_{-}^{13})},  \label{resz1}
\end{eqnarray}%
where $\epsilon _{\pm }^{ij}\equiv \epsilon _{i}\pm \epsilon _{j}$. The last
two ratios are ill-defined distributions. We want to show that their
contributions drop out. When $\epsilon \rightarrow 0$, the identity%
\begin{equation*}
\frac{1}{\omega _{1}-\omega _{2}-p^{0}-i\epsilon _{-}^{12}}=\mathcal{P}\frac{%
1}{\omega _{1}-\omega _{2}-p^{0}}+i\pi \mathrm{sgn}(\epsilon
_{-}^{12})\delta (\omega _{1}-\omega _{2}-p^{0}),
\end{equation*}%
tells us that the ill-defined part is the one proportional to $\mathrm{sgn}%
(\epsilon _{-}^{12})$. It is easy to check that in the expression (\ref%
{resz1}) sgn$(\epsilon _{-}^{12})$ multiplies%
\begin{eqnarray}
&&\frac{i\pi M^{12}}{4\omega _{1}\omega
_{2}(m_{1}^{4}+M^{4})(m_{2}^{4}+M^{4})}\frac{1}{|(\omega _{1}-q^{0})^{2}-\nu
_{3}^{2}|^{2}}  \notag \\
\qquad  &&\qquad \times \frac{1}{(\omega _{1}+\omega _{3}-q^{0}-i\epsilon
_{+}^{13})}\frac{1}{(\omega _{1}-\omega _{3}-q^{0}-i\epsilon _{-}^{13})}%
\delta (\omega _{1}-\omega _{2}-p^{0})  \label{fist}
\end{eqnarray}%
and cancels an analogous contribution coming from Res$(z_{2})$, which can be
obtained by exchanging the poles $z_{1}$ and $z_{2}$, i.e. $\omega _{1}$
with $\omega _{2}+p^{0}$, as well as $\epsilon _{1}$ with $\epsilon _{2}$.
Since (\ref{fist}) is invariant under these operations, but sign($\epsilon
_{-}^{12}$) turns into its opposite, the total vanishes.

A similar contribution to Res$(z_{1})$, proportional to sgn$(\epsilon
_{-}^{13})$ cancels a contribution due to Res$(z_{3})$. Formula (\ref{resz1}%
) also contains a term equal to sgn$(\epsilon _{-}^{12})$sgn$(\epsilon
_{-}^{13})$ times%
\begin{equation}
-\frac{\pi ^{2}M^{12}\delta (\omega _{1}-\omega _{2}-p^{0})\delta (\omega
_{1}-\omega _{3}-q^{0})}{8\omega _{1}\omega _{2}\omega
_{3}(m_{1}^{4}+M^{4})(m_{2}^{4}+M^{4})(m_{3}^{4}+M^{4})}.  \label{bin}
\end{equation}%
Summing the contributions of this type due to the three standard residues $%
z_{1}$, $z_{2}$, $z_{3}$, and noting that%
\begin{equation*}
\text{sgn}(\epsilon _{-}^{12})\text{sgn}(\epsilon _{-}^{13})+\text{sgn}%
(\epsilon _{-}^{23})\text{sgn}(\epsilon _{-}^{21})+\text{sgn}(\epsilon
_{-}^{31})\text{sgn}(\epsilon _{-}^{32})=1,
\end{equation*}%
which is easy to prove by choosing $\epsilon _{1}>\epsilon _{2}>\epsilon
_{3} $, the total is (\ref{bin}), which has no imaginary part. We obtain a
purely imaginary contribution to $i\mathcal{M}$ (an $i$ factor being brought
by the residue theorem). The contributions of this type drop out of the
cutting equation, whose left-hand side $i\mathcal{M}-i\mathcal{M}^{\ast }$
is manifestly real, if we manage to write the right-hand side in a
manifestly real form.

\subsection*{B. Unitarity}

Collecting the results found so far, we get 
\begin{equation}
i\mathcal{M}-i\mathcal{M}^{\ast }=2\lambda ^{3}\int \frac{\mathrm{d}^{D-1}%
\mathbf{k}}{(2\pi )^{D-1}}\mathrm{Im}\left[ \mathrm{Res}(z_{1})+\mathrm{Res}%
(z_{2})+\mathrm{Res}(z_{3})\right] .  \label{discm}
\end{equation}%
Dropping the ill-defined distributions, Res$(z_{1})$ effectively simplifies
to%
\begin{eqnarray*}
&&\frac{M^{12}}{2\omega _{1}(m_{1}^{4}+M^{4})}\frac{1}{|(\omega
_{1}-p^{0})^{2}-\nu _{2}^{2}|^{2}|(\omega _{1}-q^{0})^{2}-\nu _{3}^{2}|^{2}}%
\mathcal{P}\frac{1}{\omega _{1}-\omega _{2}-p^{0}}\mathcal{P}\frac{1}{\omega
_{1}-\omega _{3}-q^{0}} \\
&&\qquad \qquad \qquad \qquad \times \frac{1}{(\omega _{1}+\omega
_{2}-p^{0}-i\epsilon _{+}^{12})(\omega _{1}+\omega _{3}-q^{0}-i\epsilon
_{+}^{13})}.
\end{eqnarray*}%
Now, observe that all the ratios that appear here are real except the last
one, which has the form 
\begin{equation*}
\frac{1}{a-i\epsilon }\frac{1}{b-i\epsilon ^{\prime }}.
\end{equation*}

We need to calculate the imaginary part of this expression, which can be
handled by using the identity%
\begin{eqnarray}
\mathrm{Im}\left[ \frac{1}{a-i\epsilon }\frac{1}{b-i\epsilon ^{\prime }}%
\right] &=&\frac{1}{2i}\left( \frac{1}{a-i\epsilon }-\frac{1}{a+i\epsilon }%
\right) \frac{1}{b+i\epsilon ^{\prime }}+\frac{1}{2i}\left( \frac{1}{%
b-i\epsilon ^{\prime }}-\frac{1}{b+i\epsilon ^{\prime }}\right) \frac{1}{%
a-i\epsilon }  \notag \\
&=&\frac{\pi \delta (a)}{b+i\epsilon ^{\prime }}+\frac{\pi \delta (b)}{%
a-i\epsilon }.  \label{ima}
\end{eqnarray}%
The first contribution of the last line leads to%
\begin{equation}
-\frac{\pi \delta (\omega _{1}+\omega _{3}-q^{0})}{4\omega _{1}\omega
_{3}(m_{1}^{4}+M^{4})(m_{3}^{4}+M^{4})}\frac{M^{12}}{|(\omega
_{1}-p^{0})^{2}-\nu _{2}^{2}|^{2}}\frac{1}{\omega _{1}+\omega
_{2}-p^{0}+i\epsilon _{+}^{12}}\mathcal{P}\frac{1}{\omega _{1}-\omega
_{2}-p^{0}}.  \label{appe}
\end{equation}%
Now, observe that if we replace the principal value in this expression with
another prescription, the difference%
\begin{equation}
-\frac{i\pi ^{2}M^{12}\delta (\omega _{1}+\omega _{3}-q^{0})\delta (\omega
_{1}-\omega _{2}-p^{0})}{8\omega _{1}\omega _{2}\omega
_{3}(m_{1}^{4}+M^{4})(m_{2}^{4}+M^{4})(m_{3}^{4}+M^{4})}  \label{diffe}
\end{equation}%
is purely imaginary. The contributions of this type cancel out from the
formula for $i\mathcal{M}-i\mathcal{M}^{\ast }=i\mathrm{Disc}\mathcal{M}$,
as long as we manage to write it in a manifestly real way. We proceed by
changing the prescription in a convenient way and check the cancelations in
the final result (\ref{cuttra}).

\begin{figure}[t]
\begin{center}
\includegraphics[width=8cm]{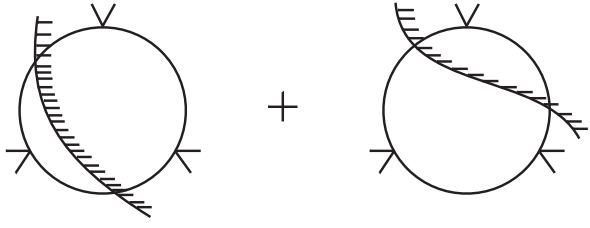}
\end{center}
\caption{Cut triangle diagrams}
\label{trianglecut}
\end{figure}

With a new prescription, the contribution of (\ref{appe}) to $i\mathcal{M}-i%
\mathcal{M}^{\ast }$ can be turned into 
\begin{equation}
-\lambda ^{3}\int \frac{\mathrm{d}^{D}k}{(2\pi )^{D}}\theta
(k^{0})D_{c}(k^{2},m_{1}^{2})D^{\ast }((k-p)^{2},m_{2}^{2},\epsilon
_{+}^{12})\theta (q^{0}-k^{0})D_{c}((k-q)^{2},m_{3}^{2})\equiv -\mathcal{C}%
_{1},  \label{gut2}
\end{equation}%
where $\mathcal{C}_{1}$ is the first cut diagram of fig. \ref{trianglecut},
calculated with the right cut propagators [i.e. $\theta (\pm
p^{0})D_{c}(p^{2},m^{2})$], which propagate only the physical degrees of
freedom. Similarly, the second contribution of (\ref{ima}) gives $-\mathcal{C%
}_{2}$, where $\mathcal{C}_{2}$ is the second cut diagram.

Repeating the same steps for $z_{2}$ and $z_{3}$ we find minus the other
four cut diagrams $\mathcal{C}_{j}$, $j=3,\ldots 6$, which can be obtained
by permuting the vertices of the cut diagrams shown in fig. \ref{trianglecut}%
. The total gives the correct cutting equation 
\begin{equation}
i\mathcal{M}-i\mathcal{M}^{\ast }=-\sum_{j=1}^{6}\mathcal{C}_{j}.
\label{cuttra}
\end{equation}%
Note that the right-hand side of this formula is manifestly real, as
promised. Indeed, its imaginary part is obtained by replacing the noncut
propagators with delta functions, which gives contributions evaluated on the
physical poles of all three propagators. This makes them equal to analogous
contributions found in the cutting equation of the standard triangle, times
a real constant. Since those contributions cancel out in the case of the
standard triangle, the right-hand side of (\ref{cuttra}) is also real.

Again, formula (\ref{cuttra}), which is nothing but the identity $%
iT-iT^{\dag }=-TT^{\dag }$ in the particular case of the triangle diagram,
shows that no unphysical degrees of freedom propagate through the cuts,
which confirms the perturbative unitarity of the LW model.

\section{Unitarity with nontrivial numerators}

\label{nontrinum}

So far, we have considered only theories with nonderivative vertices. For
most applications, such as quantum gravity and gauge theories, it is
necessary to include the case where vertices carry derivatives, which leads
to nontrivial numerators in the loop integrals. We show that their presence
does not change the previous results. We understand that the dimensional
regularization is used, which makes it possible to apply the residue theorem
even when the integral on the energy is divergent (for details on this, see
the appendix of ref. \cite{ugo}).

We assume that the Lagrangian is Hermitian, because this is an essential
requirement for unitarity. Then the vertices, which carry an additional
factor $i$, are anti-Hermitean. Denote a vertex with $n$ legs by $V_{\alpha
_{1}\ldots \alpha _{n}}^{\mu _{1}\ldots \mu _{n}}(p,k)$, where $p$ and $k$
denote the external and loop momenta, respectively, $\mu _{i}$ are Lorentz
indices and $\alpha _{i}$ are any other indices. We can decompose it as%
\begin{equation*}
V_{\alpha _{1}\ldots \alpha _{n}}^{\mu _{1}\ldots \mu
_{n}}(p,k)=\sum_{j}A_{\alpha _{1}\ldots \alpha _{n}}^{(j)}T_{j}^{\mu
_{1}\ldots \mu _{n}}(p,k),
\end{equation*}%
where $T_{j}^{\mu _{1}\ldots \mu _{n}}$ are real tensor polynomials and $%
A_{\alpha _{1}\ldots \alpha _{n}}^{(j)}$ are constant anti-Hermitian
tensors. We can focus on loop integrals with numerators made of products of
tensors $T_{j}^{\mu _{1}\ldots \mu _{n}}(p,k)$.

Unitarity still holds, since the main arguments of the previous sections are
determined by the locations of the poles, which do not change. For example,
let us check that the ill-defined distributions cancel out. Consider
formulas (\ref{resi}). If a numerator is present, we can incorporate it into
the function $h$ that appears in the sum $\mathrm{Res}(z_{1})+\mathrm{Res}%
(z_{2})$ of (\ref{resa}), so the total (\ref{total}) is still regular. The
argument is basically the same for the triangle and more complicated
diagrams. Moreover, the crucial identities (\ref{ress}) and (\ref{crucial})
still hold. The Hermiticity of the Lagrangian ensures (\ref{crucial}) up to
the effects due to $i\epsilon $, which still have the form shown on the
right-hand side of (\ref{ress}). Specifically, the limit $\epsilon
\rightarrow 0$ generates integrable singularities $\sim 1/z$, where $z$ is a
complex function of the integration variables that cannot vanish when the LW
pinching takes place. Then, the two contributions of $\tilde{\delta}(z)$
cancel each other. In the end, the residues of the LW poles simplify in the
cutting equations, so the only contributions that survive are those coming
from the standard poles. It is also clear that most of these features are
independent of the particular diagram that we are considering, so we expect
them to hold in every diagram.

\section{Violations of unitarity in Minkowski higher-derivative theories}

\setcounter{equation}{0}\label{Minkviol}

The LW models are defined as nonanalytically Wick rotated Euclidean
higher-derivative theories. In sections \ref{LWbubble} and \ref{LWuntriangle}
we have made explicit calculations to verify that they satisfy the unitarity
equation.

It is interesting to inquire whether the Minkowski versions of the same
models satisfy the unitarity equation or not. By \textquotedblleft Minkowski
versions\textquotedblright\ we mean that the integrals on the energies are
not performed along the integration path of fig. \ref{wick} derived in
section \ref{LWE}, but along the real $k^{0}$ axis, as we would normally do.
The integration path splits each LW pair into a pole above the path and a
pole below the path. We expect that the unitarity equation is violated in
this case. However, the violation is not visible at the tree level, because
no energy integrals are involved: the tree cutting equations of the
Minkowski models are identical to those of the LW models. Thus, it is
necessary to make a one-loop calculation.

In this section we prove that the bubble diagram of the Minkowski models
does not satisfy the expected cutting equation. The example we consider is a
particular case where the nonlocal divergences pointed out in ref. \cite{ugo}
are absent.

The loop integral is still (\ref{bubblo}), but now, when we integrate on $%
k^{0}$ and close the integration path in the lower half plane, we get
contributions from a different set of residues. We still have the physical
poles $z_{1}$ and $z_{2}$ of (\ref{reso}), as well as $w_{1}^{\ast }$ and $%
w_{2}^{\ast }$. However, we have $w_{1}^{\prime }=-\nu _{1}$ and $%
w_{2}^{\prime }=p^{0}-\nu _{2}$ instead of $w_{1}$ and $w_{2}$.

If the Minkowski theory were unitary, its cut diagrams would coincide with
those of the LW theory, because the physical degrees of freedom are the
same. Thus, if we subtract the cutting equation (\ref{bubblecutf}) of the LW
theory from the one of the Minkowskian theory, the right-hand side should
give zero. We show that, instead, the discontinuity $\mathrm{Disc}\left( 
\mathcal{M}_{M}-\mathcal{M}_{E}\right) $ of the difference $\mathcal{M}_{M}-%
\mathcal{M}_{E}$ between the Minkowski amplitude $\mathcal{M}_{M}$ and the
nonanalytically Wick rotated Euclidean amplitude $\mathcal{M}_{E}$ does not
vanish.

\begin{figure}[t]
\begin{center}
\includegraphics[width=8truecm]{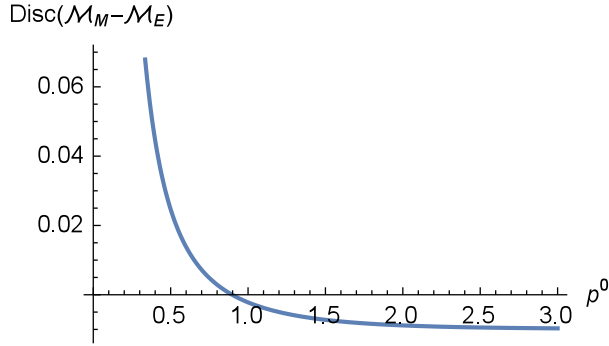}
\end{center}
\caption{Difference between the Minkowski and Wick rotated Euclidean
theories }
\label{Mink}
\end{figure}

Specifically, we find%
\begin{equation*}
\mathrm{Disc}\left( \mathcal{M}_{M}-\mathcal{M}_{E}\right) =i\lambda
^{2}\int \frac{\mathrm{d}^{D-1}\mathbf{k}}{(2\pi )^{D-1}}\mathrm{Im}\left[ 
\text{\textrm{Res}}(\nu _{1})+\text{\textrm{Res}}(p^{0}+\nu _{2})-\text{%
\textrm{Res}}(-\nu _{1})-\text{\textrm{Res}}(p^{0}-\nu _{2})\right] .
\end{equation*}%
We can simplify the calculation by choosing $\mathbf{p}=0$, $m_{1}=m_{2}=0$.
Then we have $\nu _{1}=\nu _{2}=\sqrt{\mathbf{k}^{2}+iM^{2}}$, $\omega
_{1}=\omega _{2}=|\mathbf{k}|$. Setting $M=1$, the behavior of $\mathrm{Disc}%
\left( \mathcal{M}_{M}-\mathcal{M}_{E}\right) $ as a function of $p^{0}$ is
nontrivial. Numerically, we find the plot of fig. \ref{Mink} times $i\lambda
^{2}$. This proves that the Minkowski theories violate perturbative
unitarity.

\section{Conclusions}

\setcounter{equation}{0}\label{conclu}

In this paper, we have investigated the perturbative unitarity of the
Lee-Wick models, formulated as nonanalytically Wick rotated Euclidean
theories. We have shown that it is possible to study the cutting equations
in each analytic region $\mathcal{A}_{i}$ of the complex plane, by deriving
them in suitable subdomains $\mathcal{O}_{i}$ and then analytically
extending the equations to the whole regions $\mathcal{A}_{i}$. The unitary
cutting equations hold in each $\mathcal{A}_{i}$, with no propagation of
unphysical degrees of freedom. We have made explicit computations in the
cases of the bubble and triangle diagrams, but the derivations can be
extended to all the one-loop diagrams. Moreover, the basic arguments do not
appear to depend on the specific cases we have dealt with, so we believe
that the conclusions hold for all diagrams.

On the other hand, the Minkowski versions of the same higher-derivative
theories violate unitarity. In a way or another (violations of the locality
of counterterms as shown in ref. \cite{ugo}, or violations of unitarity) the
Minkowski higher-derivative theories are not viable. This means that, in
some sense, quantum field theory prefers what we may call \textquotedblleft
Wick spacetime\textquotedblright , i.e. the Wick rotated Euclidean space, to
the Minkowski spacetime. \vskip 12truept \noindent {\large \textbf{%
Acknowledgments}}

\vskip 12truept

We are grateful to U. Aglietti and L. Bracci for useful discussions.

\end{document}